\documentclass[twocolumn,tighten,times]{aastex62}
\usepackage{color}
\usepackage{multirow}
\usepackage{chngcntr}
\usepackage{mathtools}
\usepackage{amsmath}
\usepackage{amsfonts}
\usepackage{amssymb}
\usepackage[utf8]{inputenc}  % 使用标准的 utf8 编码
\usepackage{graphicx}        % 确保这里是打开的
\usepackage{perpage}         % 在脚注中使用
\usepackage{bm}
\MakePerPage{footnote}

\graphicspath{{./}{figures/}}

\makeatletter

\newcommand{\Rmnum}[1]{\expandafter\@slowromancap\romannumeral #1@}
\makeatother

\received{\today}
\revised{tomorrow}
\accepted{the day after tomorrow}

\submitjournal{ApJ Letters}

\shorttitle{AM alignment of cluster}
\shortauthors{Rong et al.}

\begin{document}

\title{Orthogonal Alignment of Galaxy Group Angular Momentum with Cosmic Filament Spines: An Observational Study}

\correspondingauthor{Yu Rong; Peng Wang}
\email{rongyua@ustc.edu.cn; pwang@shao.ac.cn}

\author[0000-0002-2204-6558]{Yu Rong}
\affiliation{Department of Astronomy, University of Science and Technology of China, Hefei, Anhui 230026, People's Republic of China}
\affiliation{School of Astronomy and Space Sciences, University of Science and Technology of China, Hefei 230026, Anhui, People's Republic of China}

\author[0000-0003-2504-3835]{Peng Wang}
\affiliation{Shanghai Astronomical Observatory, Chinese Academy of Sciences, Nandan Road 80, Shanghai 200030, People's Republic of China}

\author[0009-0001-7527-4116]{Xiao-xiao Tang}
\affil{Shanghai Astronomical Observatory, Chinese Academy of Sciences, Nandan Road 80, Shanghai 200030, People's Republic of China}
\affil{University of Chinese Academy of Sciences, Beijing 100049, China}

%%%%%%%%%%%%%%%%%%%%%%%%%%%%%%%%%%%%%%%%
\begin{abstract}
	
	We investigated the alignment between the angular momenta of galaxy groups and the spines of their associated cosmic filaments. Our results demonstrate a significant tendency for these two orientations to be perpendicular, indicating that the rotation of a galaxy group does not originate from the spin of cosmic filaments. Instead, it is driven by the orbital angular momentum contributed by member galaxies as they accrete along the direction of the filament spines. Moreover, the strength of this perpendicular alignment signal varies with the richness of the galaxy groups, with the most pronounced alignment observed among the wealthiest groups. This pronounced alignment is largely due to the more coherent spatial distribution of member galaxies in richer groups relative to the filament spines. Our study provides valuable insights into the mechanisms of angular momentum acquisition in galaxy groups from an observational standpoint.

\end{abstract}

\keywords{galaxies: formation --- galaxies: evolution --- methods: statistical --- galaxies: photometry}

%%%%%%%%%%%%%%%%%%%%%%%%%%%%%%%%%%%%%%%%%

\section{Introduction}           %% first-level sections will be auto-capitalized
\label{sec:1}

The hierarchical genesis of cosmic structures, ranging from small-scale galaxy pairs to large-scale galaxy clusters and groups, constitutes a fundamental aspect of our understanding of cosmic evolution. According to the hierarchical evolution theory of cosmic structures, galaxy clusters and groups gradually form through the amalgamation of individual galaxies originating from large-scale filaments. Investigating these cosmic entities yields invaluable insights into both the formation and evolution of the galaxies they encompass and the large-scale structures themselves.

Galaxy clusters and groups represent the largest virialized structures in the universe, containing hundreds to thousands of galaxies. A pivotal aspect of these structures is their angular momentum properties. The angular momentum of a galaxy cluster or group, which encapsulates the collective rotational motion of its constituent galaxies, serves as a crucial parameter for elucidating the complex dynamics and evolutionary trajectories of these cosmic structures. A pertinent question arises regarding the origins of the angular momentum in galaxy groups and clusters. Recent observational and simulation studies \citep{2021NatAs...5..839W, Xia21, Sheng22, Tang25, Wang25} indicate that large-scale filaments exhibit self-rotation, with their rotation axes aligned with their spines (see Fig.~\ref{sketch}). As nodes at which filaments converge, galaxy clusters and groups may inherit the rotational angular momentum direction from these filaments, suggesting that the angular momentum direction of galaxy clusters and groups could be in principle aligned parallel to the direction of the filament spines.

Conversely, investigations into galaxy pairs and triplets reveal that their systemic angular momentum tends to be perpendicular to the direction of the filament spines, attributed to the distribution of member galaxies along these structures \citep{Tempel15b,Rong24}. For massive galaxy clusters and groups, given that their member galaxies predominantly accrete along the filaments, one might expect a similar distribution along the direction of the filament spines \citep[e.g.,][]{Knebe04}. However, due to the large number of member galaxies, the overall angular momentum direction of the system may exhibit greater complexity, and could potentially lack a clear correlation with the filament direction.

%Alignment analysis provides a potent methodology for scrutinizing and constraining scenarios of structure formation. Previous studies have investigated the spatial alignment of small galaxy groups, such as galaxy pairs and galaxy triplets, and 
A systematic investigation into the angular momentum alignment within virialized systems of varying scales, in relation to their associated filaments, yields essential insights into the fundamental mechanisms that govern their formation and evolution. Over the past decade, the relationship between the angular momentum orientations of galaxies, which represent small virialized units, and cosmic filaments has been extensively studied. Cosmological simulations have demonstrated that the angular momentum of low-mass galaxy halos tends to align their spin directions parallel to the filament spines, likely due to anisotropic collapse and tidal torques \citep{Codis12,Laigle15,Libeskind13,Ganeshaiah19,Barsanti22,Kraljic21,Tempel13a,Tempel13b,Zhang15,Dubois14,WangKang2017,WangKang2018}. In contrast, the angular momentum of high-mass halos is more frequently oriented perpendicular to the filament spines. This phenomenon arises because these systems form primarily through galaxy mergers, which predominantly occur along the direction of the filament spines due to the spatial distribution of galaxies along these structures \citep{Lee15,Tempel15a,WangKang2018}. As a result, the orbital angular momentum associated with these mergers tends to be perpendicular to the filament spines; Following the merger, this orbital angular momentum is transformed into the rotational angular momentum of the resulting halo, thus aligning it perpendicular to the filament spine \citep{Tempel13a,Tempel13b,Tempel15a,Barsanti22,Kraljic21,Kitzbichler03,Rong15b,Rong16,Rong24b}.

This study aims to investigate the alignment of angular momentum in galaxy groups within cosmic filaments in observations. Section~\ref{sec:2} outlines the sample utilized in this investigation. Section~\ref{sec:3} presents the methodology for estimating the rotation axes of galaxy groups as viewed on the celestial plane. Section~\ref{sec:4} analyzes the alignment between the angular momentum vectors and filament spines on the celestial plane. Finally, Sections~\ref{sec:5} and \ref{sec:6} provide a discussion and summary of the findings, respectively. Throughout this paper, we assume the seven-year Wilkinson Microwave Anisotropy Probe (WMAP) cosmology: the Hubble constant $H_0 = 100\ h\ {\rm{km\ s^{-1}\ Mpc^{-1}}}$, $h=0.70$, matter density $\Omega_{\rm{m}} = 0.27$, and dark energy density $\Omega_{\rm{\Lambda}} = 0.73$ \citep{Komatsu11}.

%%%%%%%%%%%%%%%%%%%%%%%%%%%
%\section{Observation}
%\label{sec:2}
\section{Sample}
\label{sec:2}

We employ the spectroscopic galaxy sample curated by \cite{Tempel12}, which is based on data from the Sloan Digital Sky Survey \citep[SDSS;][]{Aihara11}. This sample is confined to an $r$-band Petrosian magnitude threshold of $m_r = 17.77$~mag and encompasses a variety of properties for each galaxy, including its spatial coordinates, $ugriz$-band absolute magnitudes, group ranking (`rank'), and other pertinent characteristics. The identification of galaxy groups and clusters was accomplished in the work of \cite{Tempel12} using the Friends-of-Friends (FOF) algorithm, which yielded estimates of the richness and virial radius for each galaxy group or cluster.

For every galaxy member within a group, the stellar mass, $M_{\star}$, is estimated from the $r$-band absolute magnitude and $g-r$ color, utilizing the mass-to-light ratio described by $\log(M_{\star}/L_r) = 1.097(g-r)-0.306$ \citep{Bell03}. This method for estimating stellar mass based on color carries a typical uncertainty of approximately 0.10\--0.15~dex. In this study, we designate galaxies with `rank=1', i.e. the brightest members, as the centers of their respective groups. We have verified that using the most massive members as group centers would not alter any of our conclusions.

In a Cartesian coordinate system, the position of the $i$-th member galaxy in a group within a group on the celestial plane is determined using the methodology delineated by \cite{Tempel14a}, expressed as,
\begin{equation}
\begin{aligned}
    & x_i = -\sin \lambda_i, \\
    & y_i = \cos \lambda_i \cos \eta_i, \\
    & z_i = \cos \lambda_i \sin \eta_i, \\
    & \bm{r}_i=[x_i,y_i,z_i]-[x_{\rm c},y_{\rm c},z_{\rm c}],
\end{aligned}
\end{equation}
where $\lambda_i$ and $\eta_i$ denote the SDSS survey coordinates of the $i$-th member galaxy. The vector $\bm{e}_{\rm{c}}=[x_{\rm c},y_{\rm c},z_{\rm c}]$ represents the position vector of the group center, while $\bm{r}_i$ describes the position vector of the $i$-th member relative to the group center on the celestial plane.

The filament associated with each galaxy group is identified using the filament catalog compiled by \cite{Tempel14a}, based on the three-dimensional distance $d_{\rm{gf}}$ from the center of the group to the spine of the filament. The alignment analysis is confined to groups for which $d_{\rm{gf}}\leq 1.0$~Mpc$/h$, which approximately delineates a filament boundary \citep{Wang24}. This final sample comprises 30,880 galaxy groups situated within their corresponding host filaments.

The three-dimensional (3D) orientation of the spine of the filament, associated with each galaxy group, is provided by \cite{Tempel14a}. The orientation of the nearest filament point
%to the center of the group 
is adopted as the orientation of the filament spine, denoted as $\bm{e}_{\rm f}=[{\rm{d}}x,{\rm{d}}y,{\rm{d}}z]$. For further details on the methodology for estimating the 3D orientations of cosmic filaments, we direct the reader to \cite{Tempel14a}.

%%%%%%%%%%%%%%%%%%%%%%%%%%%%%%%%%%%%%%

\section{Group angular momentum}
\label{sec:3}

In this study, we investigate the potential alignment between the angular momenta of galaxy groups, denoted as $\bm{L}$, and the spine orientations of their parent cosmic filaments, represented as $\bm{e}_{\rm f}$. We note that, if the two vectors exhibit parallelism or orthogonality in 3D space, we expect to observe corresponding signals of parallelism or orthogonality on the two-dimensional (2D) celestial plane. 

To mitigate potential systematic biases introduced by redshift-based distance measurements, we restrict our analysis to alignment signals within the celestial plane. This approach is motivated by the significant impact of the ``Fingers of God'' effect, a redshift-space distortion phenomenon caused by the peculiar velocities of galaxies within groups and filaments. These velocities induce elongated distortions along the line-of-sight in redshift space, thereby introducing systematic errors in the inferred distances of individual galaxies. Given that angular momentum is derived from the spatial distribution and velocities of member galaxies, errors in their inferred distances propagate into the angular momentum calculation, potentially misaligning it relative to the true physical orientation. Consequently, the alignment between group angular momentum and large-scale filament orientations may be systematically influenced by these redshift-space distortions. Through this methodological restriction to the celestial plane, we aim to minimize these systematic effects and obtain more reliable measurements of alignment signals in the large-scale structure of the universe.

The filament orientation vector $\bm{e}_{\rm f}$ is projected onto the celestial plane as,
\begin{equation}
\begin{aligned}
    &\bm{e}_{\rm f}'=\bm{e}_{\rm f}-(\bm{e}_{\rm f}\cdot \bm{e}_{\rm c})\bm{e}_{\rm c},\\
    &\bm{e}_{\rm f}'=\bm{e}_{\rm f}'/|\bm{e}_{\rm f}'|.
\end{aligned}
\end{equation}

We then define the component of $\bm{L}$ in the celestial plane, $\bm{L}_{\rm p}$, as
\begin{equation}
\begin{aligned}
	& \bm{L}_{\rm p}\equiv \Sigma_{i=1}^{N} M_i\bm{r}_i\times \bm{v}_{i},\\
	& \bm{e}_{L\rm p}=\bm{L}_{\rm p}/|\bm{L}_{\rm p}|,
\label{am}
\end{aligned}
\end{equation}
where $N$ denotes the richness of a group, $M_i$ represents the mass weight of the $i$-th member galaxy, $\bm{v}_i$ indicates the relative radial velocity vector of the $i$-th member galaxy along the line of sight, defined as 
\begin{equation}
    \bm{v}_i = (v_i-v_{\rm c})[x_i,y_i,z_i],
\end{equation}
with $v_i$ and $v_{\rm c}$ denoting the radial velocities of the $i$-th member galaxy and the brightest galaxy in the group, respectively. These velocities are derived from the spectroscopically observed redshifts $z_{\rm{obs}}$, according to the relation $v\simeq cz_{\rm{obs}}$, where $c$ is the speed of light.

The observational quantity $\bm{L}_{\rm p}$ is perpendicular to the line-of-sight, thereby facilitating the characterization of the rotational motion of a galaxy group along this axis; it effectively represents the rotation axis of the group within the celestial plane. In Section~\ref{sec:5}, we will elucidate that if $\bm{L}$ and $\bm{e}_{\rm f}$ exhibit parallelism or orthogonality in the 3D space, $\bm{e}_{L\rm p}$ and $\bm{e}_{\rm f}'$ similarly exhibit parallelism or orthogonality in the 2D projection.

%%%%%%%%%%%%%%%%%%%%%%%%%%%%%
\section{Alignment of group angular momentum with filament}
\label{sec:4}

Here, we assume the mass weight $M_i=M_{\star}$. Subsequently, the intersection angle $\delta$ between $\bm{e}_{L\rm p}$ of each group and the projected filament vector $\bm{e}_{\rm f}'$ is computed as
\begin{equation}
   \delta={\rm acos} (\bm{e}_{L\rm p}\cdot \bm{e}_{\rm f}').
\end{equation}
The angle $\delta$ is constrained within the interval $[0,90^{\circ}]$.

In panel~a of Fig.~\ref{fig1}, we illustrate the distribution of $\delta$ for galaxy groups categorized by richness: groups with rich memberships (i.e., $N>10$; 1,356 groups), those of intermediate richness (i.e., $3<N\leq 10$; 8,557 groups), and groups with poor memberships (i.e., $2\leq N\leq 3$; 20,967 groups). The distributions of $\delta$ for these three subsamples display a significant deviation from a uniform distribution, evidenced by Kolmogorov–Smirnov (K-S) test (compared to a uniform distribution) $p$-values of $10^{-18}$, $10^{-54}$, and $10^{-70}$,  respectively. 

Following the methodology delineated in \cite[see also \cite{Rong19,Rong24}]{Rong24b}, we define the index ${\mathcal{I}}(\delta)=N_{0-45}/N_{45-90}$, where $N_{0-45}$ and $N_{45-90}$ signify the counts of galaxies with $\delta$ falling within the ranges [0, 45$^{\circ}$] and [45$^{\circ}$, 90$^{\circ}$], respectively. The uncertainty associated with ${\mathcal{I}}(\delta)$ is estimated via bootstrap resampling. From the original sample, consisting of $k$ members, we randomly select $k$ galaxies with replacement, iterating this procedure 1,000 times to derive 1,000 values for ${\mathcal{I}}(\delta)$. The standard deviation ($\sigma_{\mathcal{I}}$) of these values is subsequently used as a measure of uncertainty. A uniform distribution would yield ${\mathcal{I}}(\delta) \simeq 1$. Our findings reveal ${\mathcal{I}}(\delta)=0.64\pm 0.04$, $0.71\pm 0.02$, and $0.80\pm 0.01$ for the subsamples of $N>10$, $3<N\leq 10$, and $2\leq N\leq 3$, respectively. These results indicate a trend toward $\delta$ approaching $90^{\circ}$, suggesting a significant perpendicular alignment between the angular momenta of the groups and the cosmic filament spines, at a high level of confidence{\footnote{The confidence level of an alignment signal is estimated as $|1-{\mathcal{I}}(\delta)|/\sigma_{\mathcal{I}}$.}}.

Moreover, our analysis reveals the most pronounced alignment signal (i.e., the lowest ${\mathcal{I}}(\delta)$) within the subgroup characterized by rich memberships ($N>10$) members, accompanied by a noticeable reduction in the strength of this signal as group richness decreases. Given that previous studies have indicated a strong dependence of the group mass on richness \citep[e.g.,][]{Plionis06,Tovmassian09}, this trend also implies a decreasing alignment signal correlated with decreasing group mass.

\begin{figure}
   \centering
   \includegraphics[width=0.9\columnwidth, angle=0]{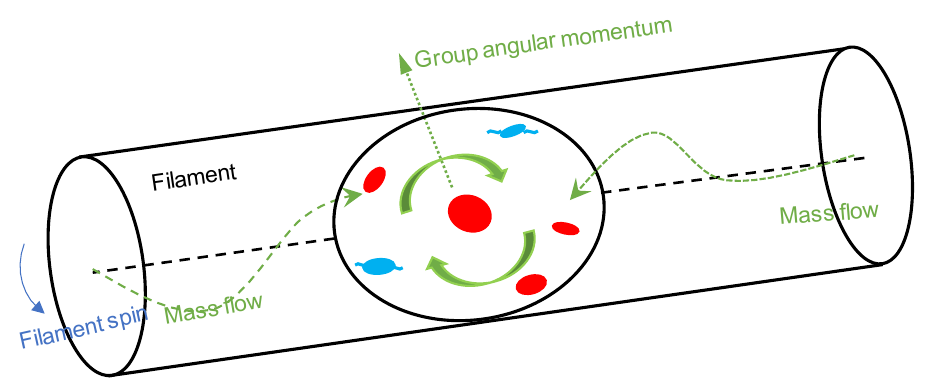}
   \caption{Schematic representation of a galaxy group and a segment of its associated large-scale filament. The galaxy group is depicted as a black ellipse, while a segment of the large-scale filament is illustrated as a black cylinder. Individual member galaxies within the group are represented by red and blue ellipses. The angular momentum of the group is indicated by dotted green arrows, and the mass flows relative to the galaxy group, signifying the infall of galaxies, are shown by dashed green arrows. The spin of the filament segment is denoted by a blue arrow.}
   \label{sketch}
   \end{figure}

\begin{figure}
   \centering
   \includegraphics[width=0.9\columnwidth, angle=0]{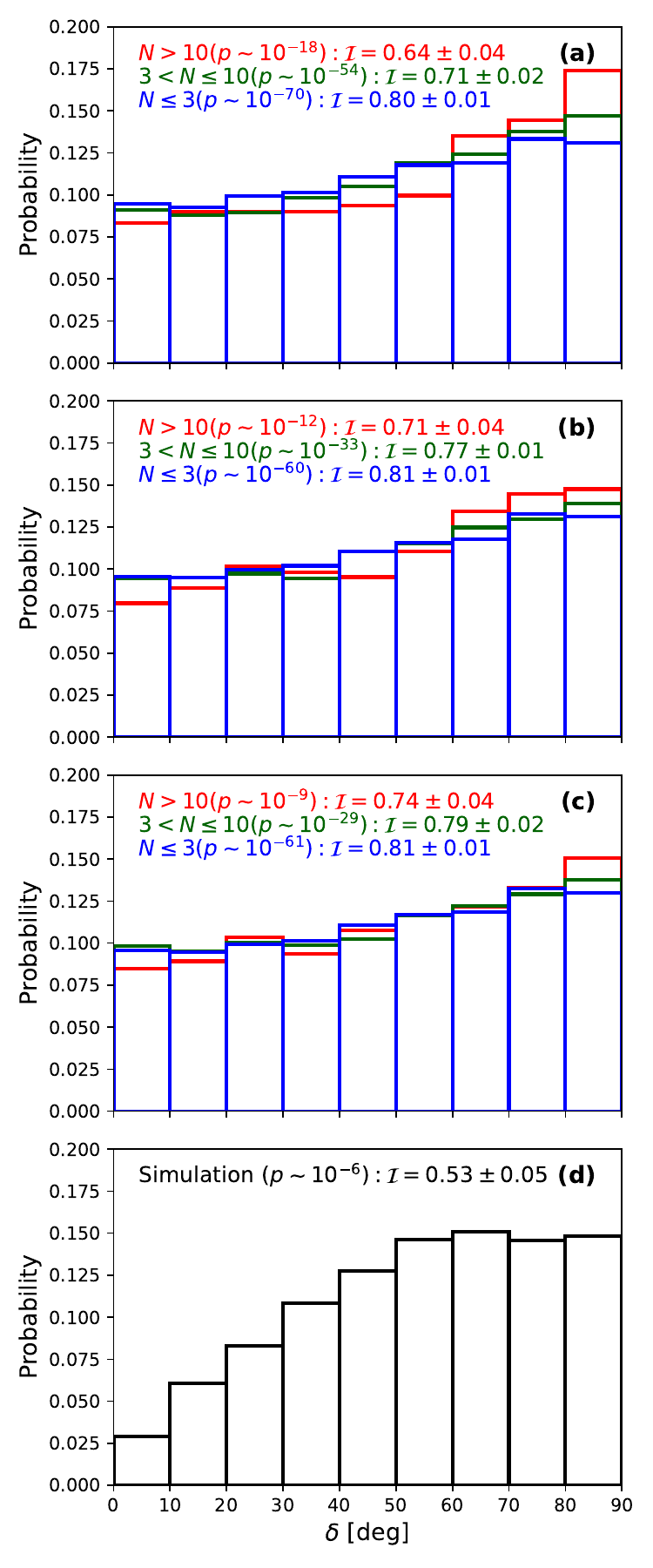}
   \caption{Panels~a, b, and c illustrate the distributions of $\delta$ for galaxy groups categorized by member richness: groups with rich members ($N>10$; red histogram), those of intermediate richness ($3<N\leq 10$; green histogram), and groups with poor members ($2\leq N\leq 3$; blue histogram). The analyses detailed in these panels employ various weighting schemes: panel~a incorporates both mass and radius weights of member galaxies, panel~b utilizes only the mass weights, and panel~c assigns equal weights to all members. Panel~d assesses the alignment in the 2D plane, evaluating whether the angular momenta of groups exhibit perpendicular alignment with the filament spines in 3D space, as derived from the Millennium-II simulation. The K-S test $p$-values of comparing the distributions and a uniform distribution are presented, alongside the values of ${\mathcal{I}}(\delta)$, highlighting the significance of the observed alignment.}
   \label{fig1}
   \end{figure}

Collectively, these observations support the hypothesis that a group's angular momentum is predominantly influenced by the infall of its constituent galaxies, rather than inheriting rotation from the filament structures. Given that galaxies generally fall into groups in alignment with the filament direction \citep[e.g.,][]{Tempel15b,Rong24}, it is expected that the direction of angular momentum would be perpendicular to the filament spine, as depicted in the accompanying diagram in Fig.~\ref{sketch}.

%%%%%%%%%%%%%%%%%%%%%%%%%

%%%%%%%%%%%%%%%%%%%%%%%%%
\section{Discussion}
\label{sec:5}

\subsection{Does the result depend on weights?}

In equation~(\ref{am}), we incorporate the mass and radius of each member galaxy as ``weights'' to determine the projected rotation axis of a galaxy group. It is important to acknowledge that galaxies situated at the periphery of a group are more likely to be misclassified as group members. In other words, galaxies with larger magnitudes of $|\bm{r}_i|$ are at an increased risk of erroneous classification \citep{Serra13}. This misidentification can lead to inaccuracies in the determination of the rotation axis in the celestial plane. Consequently, paralleling the methodology employed by \cite{Rong24a}, we also consider the application of mass weights exclusively, modifying equation~(\ref{am}) to
\begin{equation}
\begin{aligned}
	& \bm{L}'_{\rm p}\equiv \Sigma_{i=1}^{N} \frac{M_i}{|\bm{r}_i|}\bm{r}_i\times \bm{v}_{i},\\
    & \bm{e}'_{Lp}\equiv \bm{L}'_{\rm p}/|\bm{L}'_{\rm p}|.
    \label{am2}
\end{aligned}
\end{equation}
The intersection angle between the projected rotation axis and the projected filament vector is then expressed as,
\begin{equation}
   \delta'={\rm acos} (\bm{e}'_{Lp}\cdot \bm{e}_{\rm f}').
\end{equation}
As depicted in panel~b of Fig.~\ref{fig1}, we observe a consistent pattern in the results, reinforcing the conclusion that the rotational direction of the galaxy group tends to be perpendicular to the orientation of the filament spine.

Additionally, we present the alignment results without applying any weighting to the member galaxies, effectively setting $M_i=1$ for each member in equation~(\ref{am2}). As illustrated in panel c of Fig. \ref{fig1}, the perpendicular alignment signal remains clearly evident.

Upon comparing the results in panels~a, b, and c of Fig.~\ref{fig1}, we find that, irrespective of the weighting scheme employed, the angular momentum alignment signal in more richly populated groups consistently exhibits greater strength. This observation suggests that the pronounced alignment in wealthier groups may primarily stem from the spatial distributions of member galaxies being more closely aligned with the filament spine, rather than arising from variations in the weighting scheme itself. In other words, richer groups may exhibit a more elliptical morphology, a conclusion supported by earlier studies \citep[e.g.,][]{Wang08}.

%%%%%%%%%%%%%%%%%%%%%%%%%

\subsection{Does the 2D result reveal the 3D case?}

In section~\ref{sec:2}, we elucidate that if the rotation axis of a galaxy group in 3D space is perpendicular to the filament spine, then the rotation axis deduced from the 2D projection plane (celestial plane) using line-of-sight velocities will also align perpendicularly to the projected direction of the filament spine in the celestial plane. To substantiate this assertion, we quantitatively demonstrate the relationship using results from the Millennium-II $N$-body simulation \citep{Boylan-Kolchin09}. In this simulation, halos and subhalos are identified using the FOF algorithm \citep{Davis85} and the SUBFIND algorithm \citep{Springel01}, respectively.

Initially, we computed the angular momentum $\bm{L}_{\rm{s}}$ of each halo (i.e., galaxy group) at redshift $z\sim 0$ in the 3D space, utilizing the 3D velocities of the member subhalos. We restricted our analysis to halos with masses exceeding $10^{13}\ M_{\odot}/h$ and subhalos with masses greater than $10^{10}\ M_{\odot}/h$ due to limitations of simulation resolution. For the purpose of this investigation, we assume that the direction of the filament spine associated with each halo is strictly perpendicular to its angular momentum vector $\bm{L}_{\rm{s}}$, with the spine oriented arbitrarily within the plane orthogonal to $\bm{L}_{\rm{s}}$. We then project the spine onto the $y-z$ plane (with the $x$-axis representing the line-of-sight direction), designating it as $\bm{e}'_{\rm{sf}}$. Following the procedure outlined in equation~(\ref{am}), we compute the rotation axis of each halo in the $y-z$ plane, denoted as $\bm{e}_{L\rm sp}$, by considering solely the velocity components of the member subhalos along the $x$-direction.  Subsequently, we determine the angle $\delta_{\rm s}$ between $\bm{e}'_{\rm{sf}}$ and $\bm{e}_{L\rm sp}$.

As illustrated in panel~d of Fig.~\ref{fig1}, $\delta_{\rm s}$ approaches $90^{\circ}$, thereby indicating that if the angular momentum of the galaxy group in 3D space is indeed perpendicular to the direction of the filament spine, the rotation axis of the galaxy group in 2D space is likewise oriented perpendicularly to the projected filament spine. Consequently, the results derived from the 2D analysis can accurately reflect the 3D case.

%Finally, we also note that a coherent alignment of infall trajectories of member galaxies may lead to the perpendicular alignment of group angular momentum with filament. In other words, galaxies within a filament tend to have consistent angular momentum orientation when they fall into a group. Consequently, with an increasing number of infalling members, the angular momentum becomes more prominent, aligning more perpendicular to the filament spine. If the orientation of angular momentum between $\bm{L}_1$ and $\bm{L}_2$ is random when member galaxies fall into the galaxy group, some galaxies will align with $\bm{L}_1$ while others will align with $\bm{L}_2$, leading to mutual cancellation. In this scenario, the total angular momentum of a large number of member galaxies may paradoxically become weak, rather than strongly correlated with the filament spine.

%In order to examine this hypothesis, we compare the orientations of $\bm{L}$ of the different galaxy groups in the same filament. The filament catalog of \cite{Tempel14a} have provided the information for which filament points belonging to the same cosmic filament. Therefore, for each cosmic filament in the catalog of \cite{Tempel14a}, we can identify the member galaxy groups

%%%%%%%%%%%%%%%%%%%%%%%%%%%%
\section{Summary}
\label{sec:6}

Utilizing the galaxy group catalog and member galaxy data from SDSS, we investigate the alignment of the angular momenta of galaxy groups with the spines of their associated cosmic filaments as observed in the local universe. Our analysis concentrates on the alignment within the 2D celestial plane, and we examine various weighting methods to estimate the rotation axis of each group in this projection.

Our findings demonstrate that the rotation axes of galaxy groups consistently exhibit a tendency to be perpendicular to the projected filament spines, achieving a high level of statistical confidence regardless of the weighting methodologies employed. This observation suggests that, in 3D space, the angular momentum vectors of galaxy groups are likely aligned perpendicularly to their corresponding filament spines.

Moreover, the strength of the perpendicular alignment signal varies with the richness of the galaxy groups, with the most pronounced signals observed among the richest groups. For those groups containing more than 10 member galaxies, the alignment signal attains an approximate significance of $9\sigma$. The enhanced signal in wealthier groups may be attributed to the fact that the distributions of member galaxies within these groups are more closely aligned with the filament spines. In our subsequent research, we will build upon the findings of this study by utilizing simulations to further investigate the perpendicular alignment between the angular momentum of galaxy groups and their associated filament spines (Tang et al., in preparation). Specifically, we aim to elucidate the mechanisms underlying the stronger alignment signal observed in galaxy groups of greater richness through simulation-based analyses.

This study offers valuable insights into the mechanisms by which angular momentum is acquired in galaxy groups from an observational perspective: the rotation is not inherited from cosmic filaments but instead arises from the orbital angular momentum imparted by member galaxies as they fall along the direction of the filament spines. Looking forward, we intend to compare these observational results with theoretical models, including frameworks that consider cold dark matter and warm dark matter, to enhance our understanding of the dynamics of dark matter.

%%%%%%%%%%%%%%%%%%%%%%%%%%%%%%%%%%%%%%%%%%%%%%%%%%%%%%%%%%%%%%%%%%%%%%%%%%%%%%%%
%\clearpage
\acknowledgments

Y.R. acknowledges supports from the CAS Pioneer Hundred Talents Program (Category B), the NSFC grant (No. 12273037), the USTC Research Funds of the Double First-Class Initiative, and the research grants from the China Manned Space Project (the second-stage CSST science projects: ``Investigation of small-scale structures in galaxies and forecasting of observations'' and ``CSST study on specialized galaxies in ultraviolet and multi-band''). P.W. and XXT acknowledge financial support by the NSFC (No. 12473009), and also sponsored by Shanghai Rising-Star Program (No.24QA2711100).

%\end{acknowledgments}
%%%%%%%%%%%%%%%%%%%%%%%%%%%%%%%%%%%%%

%\clearpage
%%%%%%%%%%%%%%%%%%%%%%%%%%%%%%%%%%%%%
%\bibliographystyle{aastex62}

%%%%%%%%%%%%%%%%%%%%%%%%%%%%%%%%%%%%%%%%%%%%%%%%%%

\end{document}